\documentclass[prl,superscriptaddress,twocolumn,showpacs]{revtex4}
\usepackage{graphicx}
\begin{document}

\newcommand{\bi}[1]{\mbox{\boldmath ${#1}$}}
\newcommand{\sbi}[1]{\mbox{\boldmath ${\scriptstyle {#1}}$}}


\newcommand{\eref}[1]{(\ref{#1})}

\newcommand{\e}{{\rm e}}
\newcommand{\rmi}{{\rm i}}
\newcommand{\rmd}{{\rm d}}
\newcommand{\mod}{{\rm mod}}
\newcommand{\tr}{\hbox{tr}}
\newcommand{\pd}{\partial}
\newcommand{\Ree}{{\Re}{\rm e}}
\newcommand{\Imm}{{\Im}{\rm m}}
\newcommand{\half}{{\textstyle{\frac{1}{2}}}}
\newcommand{\third}{\textstyle{\frac{1}{3}}}

\newcommand{\pll}{\parallel}
\newcommand{\x}{\perp}
\newcommand{\up}{\uparrow}
\newcommand{\dn}{\downarrow}

\newcommand{\al}{\alpha}
\newcommand{\be}{\beta}
\newcommand{\de}{\delta}
\newcommand{\De}{\Delta}
\newcommand{\eps}{\epsilon}
\newcommand{\ga}{\gamma}
\newcommand{\om}{\omega}
\newcommand{\Om}{\Omega}
\newcommand{\lam}{\lambda}
\newcommand{\Lam}{\Lambda}

\newcommand{\brkt}[1]{\left({#1}\right)}
\newcommand{\sqbrkt}[1]{\left[{#1}\right]}
\newcommand{\anbrkt}[1]{\left<{#1}\right>}
\newcommand{\stbrkt}[1]{\left|{#1}\right|}
\newcommand{\cbrkt}[1]{\left\{ {#1} \right\} }
\newcommand{\bra}[1]{\left<{#1}\right|}
\newcommand{\ket}[1]{\left|{#1}\right>}

\newcommand{\ord}[1]{{\cal O} \sqbrkt{#1}}

\title{\vskip -0.4truein Berry phase in a non-isolated system} 
\author{\vskip -0.1truein Robert S. Whitney}
\affiliation{Theoretical Physics, University of Oxford, 1 Keble Road,
Oxford OX1 3NP, United Kingdom}
\author{Yuval Gefen}
\affiliation{Department of Condensed Matter Physics, 
Weizmann Institute of Science, 
Rehovot 76100, Israel}
\date{\today}
\begin{abstract}
We investigate the effect of the environment on a Berry phase
measurement involving a spin-half.  We model the spin+environment 
using a biased spin-boson Hamiltonian with a time-dependent magnetic field.
We find that, contrary to naive expectations, the Berry phase acquired by the 
spin can be observed, but only on timescales which are neither 
too short nor very long.
However this Berry phase is not the same as for the isolated spin-half.
It does not have a simple geometric interpretation in terms of 
the adiabatic evolution of either bare spin-states or the dressed 
spin-resonances that remain once we have traced out the environment.
This result is crucial for proposed Berry phase measurements in 
superconducting nanocircuits as dissipation there is known to be significant.
\end{abstract}
\maketitle

It was recently suggested \cite{Falci00} that it should be possible 
to observe the Berry phase (BP)\cite{Berry84} in a superconducting 
nanostructure, 
and possibly use it to 
control the evolution of the quantum state \cite{Jones00Ekert00,Wilhelm01}.  
This intriguing sugestion however did not consider the coupling to the 
environment, which is never negligible in such structures \cite{Nakamura99}. 
To truly understand the feasibility of the proposed experiment, 
we must know the effect of the environment on the 
BP.
Originally the BP was defined for systems whose states
were separated by finite energy gaps.  Here we ask whether a BP
can be observed in a system whose spectrum is continuous
because it is not completely isolated from its environment.
All real systems are coupled, at least weakly, to their environment and as a
result never have a truly discrete energy level spectrum.
The usual requirement for adiabaticity is that the parameters of the 
Hamiltonian 
are varied slowly compared to the gap in the spectrum.  Here there is no gap
so {\em naively} one would say that adiabaticity is impossible 
and hence the BP could never be observed.
However experiments have observed the BP, both directly and
indirectly \cite{Anandan97},
so this argument must be too naive.
We therefore take a simple model
in which a quantum system, which when isolated exhibits a BP, is coupled
to many other quantum degrees of freedom. We then ask two
questions. Firstly, under what conditions can the BP be
observed? Secondly, is the observed BP the same as that
of the isolated system? 
While others have investigated systems with a BP
coupled to other degrees of freedom
\cite{Gaitan98,Ao99,Avron98Avron99}, we believe we are the first
to explicitly address these two questions.

We distinguish between the system and the
environment in the following way.
We have complete experimental control over the {\it system}, 
but almost no control over the {\it environment}.
The most that we can do to the environment is to ensure
the ``universe'' (system $+$ environment)
is in thermal equilibrium, with a temperature $T$.
We will assume we have enough control over $T$ to take it to zero, 
and thus prepare the universe in its ground state. 
However any procedure to measure a BP in an isolated system
must involve measuring a 
phase difference from a superposition of two states.
When the system is not isolated most such procedures involve the mixing of 
a large number of eigenstates (of the universe), 
this leads to the effects that we discuss below\cite{endnote-about-Avron}.


We choose to investigate a spin-half which is coupled to both
a magnetic field and an environment (a bath of harmonic oscillators).  
Our model is a 
biased spin-boson model \cite{Leggett87} with a time-dependent field.  
When isolated from its environment,
the spin exhibits a BP if we slowly rotate the magnetic field
around a closed loop.  
This model, chosen primarily for its simplicity, is extremely relevant to
a recent proposal for observing a BP in a superconducting nanocircuit 
\cite{Falci00}.  
While we make no attempt to accurately model the true coupling between
the nanocircuit and its environment, 
we believe our results give an excellent indication of
what to expect in the real system.
Our work will also be very relevant to realisations of the
BP quantum computers proposed in \cite{Jones00Ekert00}.

In this Letter we concentrate on an {\it Ohmic environment} \cite{Leggett87},
with the universe initially at {\it zero-temperature} \cite{wg-tbp}.
We find that the spin-environment coupling causes
the spin-eigenstates to become spin-resonances which have the following 
properties.
(i) The energy distance between them is Lamb shifted by $\de E$.
(ii) The higher energy resonance exponentially decays to the lower one 
on a time-scale, $T_1$, and observables containing phase 
information exponential decay on a timescale $T_2$.
(iii) There are adiabatic phase-shifts,
which divide into two catagories with different symmetries;
the phase which vanishes when the Hamiltonian is time-independent 
we call $\de \Phi_{\rm BP}$; while those phase-shifts (and amplitudes)
which do not vanish we schematically refer to as $\Phi_{\rm shift}$.
The former scales with the winding number of the BP experiment,
while the latter does not (see below).
All of these effects go like the second power of the 
spin-environment coupling, 
see eqs. \eref{eq:results-20a}-\eref{eq:results-30}.

Effect (ii) means that one cannot perform an arbitrarily long
experiment to measure a phase:  
so we must find the BP from an
experiment where the system's Hamiltonian is taken round a closed loop in a 
finite time period, $t_{\rm p}\lesssim T_2$. 
In such an experiment
there is typically a non-zero amplitude for returning to the 
initial state and this amplitude has a phase.
We interpret the latter as the sum of a dynamic phase which
scales linearly with $t_{\rm p}$, an adiabatic phase 
($\Phi_{\rm BP}+\Phi_{\rm shift}$) 
which is independent 
of $t_{\rm p}$, and non-adiabatic contributions which are proportional 
to $\De \! \cdot \! t_{\rm p}$ to some negative power\cite{non-adiab}.  
Here $\De$ is the energy difference between the spin-resonances 
(we set $\hbar=1$).
Thus the BP is present for arbitrary 
$t_{\rm p}$, it is simply masked by the non-adiabatic contributions 
unless $t_{\rm p}$ is long enough. 
For the BP to be observed we must choose
a value for $t_{\rm p}$ which is neither too short
nor very long, so that it obeys $\De^{-1} \ll t_{\rm p} \lesssim T_2$. 
However we then actually observe a combination of 
$\Phi_{\rm BP}$ and $\Phi_{\rm shift}$.
To distinguish between these two effects we note that
when we do not rotate the Hamiltonian 
$\Phi_{\rm BP}=0$ while $\Phi_{\rm shift}$ is unchanged.

Now we ask if the environment's effect on the BP is observable.
To do this we must first decide what BP we would naively 
expect to observe.  There are two possible cases to consider:
(i) The system evolves in a magnetic field that we 
directly control, then we would expect the BP to be given by the 
solid-angle enclosed by that field, $\Phi^{(0)}_{\rm BP}$.  
The deviation from this expectation
is given by $\de \Phi_{\rm BP}$ in eq.~\eref{eq:results-20b}.
For this deviation to be observable it must be much larger than 
the non-adiabatic corrections at $t_{\rm p} \lesssim T_2$;  
this means that $\De \! \cdot \! T_2\! \cdot \! \de\Phi_{\rm BP} \gg 1$.
The functional form of $T_2$ and $\de \Phi_{\rm BP}$, 
in \eref{eq:results-20a} and \eref{eq:results-20b},  
have the same dependence on the strength of the coupling to the 
environment, $C$. 
Thus the condition reduces to one dominated by the dependence on 
$\gamma$, where $\gamma$ (defined below eq.~\eref{eq:results-20b}) 
characterises the environment.   
We conclude that there is a wide range of values of $\gamma$ for which 
we can observe $\de \Phi_{\rm BP}$.  
(ii) The second case is more complicated, but is relevant to
the superconducting nanocircuit in \cite{Falci00}.
There we have no independent measure of the bare spin Hamiltonian,
the control parameters 
(gate voltages and magnetic fluxes) enter the spin Hamiltonian in 
combination with unknown constants (capacitances and inductances).   
Thus we know nothing about the bare spin-eigenstates, or the solid angle 
that they enclose when we vary the experimental parameters.
However we can measure the spin resonances in the presence of the 
environment as a function of the experimental parameters.
Then one might predict the observed BP is given by the
solid-angle enclosed by these spin-resonances.  
This prediction is given above \eref{eq:results-40};
it is of a similar form to the correct result, but contains a very different
function of the distribution of oscillators in the environment.
The deviation from this expectation is given by $\de \Phi_{\rm BP}'$, 
for it to be observable we require that 
$\De \! \cdot \! T_2 \! \cdot \!\de \Phi_{\rm BP}' \gg 1$.
Again this reduces to a function independent of $C$,
where $\de \Phi_{\rm BP}'$ is observable over a wide range of $\gamma$.
Finally, we assume we measure $\Phi_{\rm shift}$
when for a time-independent Hamiltonian 
before carrying out the BP experiment.
Then we do not require $\de \Phi_{\rm BP}$ (or $\de \Phi_{\rm BP}'$)
to be larger than $\Phi_{\rm shift}$ for it to be observable.



\begin{figure}[t]
  \includegraphics[scale=0.9]{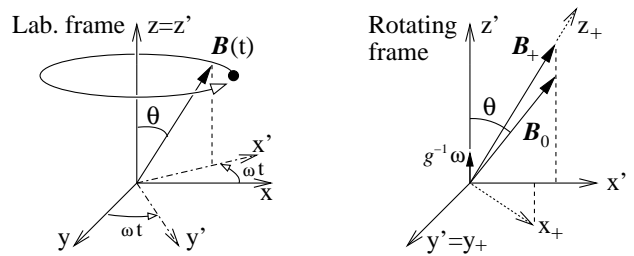} 
  \caption{Evolution in step (b) of the experiment, in Lab. and rotating 
  frames.  The primed-basis and plus-basis are both shown. 
  }
\label{fig1}
\vskip -0.2truein
\end{figure}


To be concrete we assume here that the BP is measured using 
the spin-echo method\cite{Jones00Ekert00}.
We consider an experiment where we start with the field along the $z$-axis
and the universe (spin+oscillators) is in its ground state \cite{Shnirman02}. 
(a) The field is then (instantaneously) prepared at 
its initial value, $\bi{B}_0$, which is at angle $\theta$ to the 
$z$-axis, At the same time the
spin is (intantaneously) 
placed in the state ${1 \over \sqrt{2}}\! \brkt{\ket{\up}+\ket{\dn}}$ 
relative to $\bi{B}_0$.
Then (b) we adiabatically rotate the magnetic field, $\bi{B}(t)$, 
$n$ times around a closed loop with constant angular velocity, 
$\bi{\om}= \hat{\bi{z}}\om$, (see Fig.~\ref{fig1}) for a time period, 
$t_{\rm p}=2\pi n/\om$.  We call $n$ the winding number.
After which (c) the spin is flipped and (d) the field is rotated 
with angular velocity $-\bi{\om}$ for time $t_{\rm p}$.
Finally (e) the spin is flipped again and (f) the spin state is measured.
By ``flip the spin'' we mean $\ket{\up} \leftrightarrow \ket{\dn}$,
where the $\up$ and $\dn$ are relative to the direction of 
the $\bi{B}$-field at that time.
This can be achieved by applying a instantaneous $\pi$-pulse 
oriented along the $y$-axis.  By instantaneous we mean much faster than
the fastest oscillator in the environment.
We ask what the probability is that the final spin-state, 
after carrying out (a)-(f), is in a given direction in the 
plane perpendicular to $\bi{B}_0$.
For an isolated spin-half, the 
probability of the final spin-state being 
${1\over\sqrt{2}} \! \brkt{\e^{\rmi\xi/2}\ket{\up}+\e^{-\rmi\xi/2}\ket{\dn}}$ 
is \cite{Jones00Ekert00}
\begin{eqnarray}
P(\xi) = \half 
\sqbrkt{1 + \cos\brkt{\xi - 4\Phi^{(0)}_{\rm BP}}} \ ,
\label{eq:intro-10}
\end{eqnarray}
where all spin-states are defined relative to the axis
of the field $\bi{B}_0$.  
Measuring this probability as a function of $\xi$ yields the BP 
for an isolated spin, $\Phi^{(0)}_{\rm BP}= \pi n (1-\cos \theta)$.
We wish to know what we observe if we carry out the same 
measurement for a spin which has been weakly coupled 
to a bath of oscillators throughout the experiment.


The Hamiltonian we consider
contains the spin-half in the above time-dependent magnetic field, 
$\bi{B}(t)$, which is also coupled 
to a bath of harmonic oscillators with frequencies $\{ \Om_j \}$.
Writing it in terms of creation, 
$\hat{b}^\dagger$, and annihilation, $\hat{b}$, operators 
for the oscillators,
\begin{eqnarray}
{\cal H}(t) &=& -{g \over 2} \bi{B}(t)\cdot \hat{\bi{\sigma}}
+\sum_{j,\al} \Om_j \brkt{\hat{b}^\dagger_{j,\al}\hat{b}_{j,\al} + \half}
\cr
& & -{g \over 2}\sum_{j,\al}{C_\al \over (2m\Om_j)^{1/2}} 
\brkt{\hat{b}^\dagger_{j,\al}+\hat{b}_{j,\al}}\hat{\sigma}_\al \ ,
\label{eq:details-10}
\end{eqnarray}
where $j$ is summed over all oscillators and $\al$ is summed over the 
$(x,y,z)$ components of the oscillator.  
The number of 
oscillators with frequency $\Om$ to $\Om +\rmd \Om$ is $p(\Om) \rmd \Om$.
The spectral density \cite{Leggett87} is given by 
$
J(\Om) = \sum_j \pi (gC)^2 \brkt{2m \Om_j}^{-1}\de (\Om -\Om_j)
= \pi (gC)^2 p(\Om)\brkt{2m \Om}^{-1}$.
Here we restrict ourselves to $z$-axis spin-environment coupling 
with $C_\al=C\de_{\al z}$ \cite{endnote-isotropic}.  
Then for $\bi{B}\cdot \bi{\sigma}=B \sigma_z$, 
the exact ground state of the universe\cite{Shnirman02} 
is simply $\ket{\up}\prod_j 
|0^{\scriptscriptstyle j}_{\scriptscriptstyle \up} \rangle$
where oscillator $j$ is in the ground state, 
$|0^{\scriptscriptstyle j}_{\scriptscriptstyle \up} \rangle$, of the
harmonic potential centred at $\brkt{0,0,\half g C}$.
We consider an Ohmic bath of oscillators with 
$J(\Om)= {\pi \over 2}\tilde{C}^2 \Om \exp \sqbrkt{-\Om/\Om_{\rm m}} $, 
and work in the limit of small dimensionless coupling 
$\tilde{C} \equiv  gC (\al/m)^{1/2} \ll 1$.

The time-dependence in \eref{eq:details-10} makes the problem unpleasant,
however we remove this by going to 
the primed-basis which rotates with the $\bi{B}$-field.
In this non-inertial basis the spin experiences
an effective field $\brkt{\bi{B}_0 + g^{-1}\bi{\om}}$.
For our problem the effective field is $\bi{B}_+$ for $0<t<t_{\rm p}$
(shown in Fig. \ref{fig1}), and $\bi{B}_-$ for $t_{\rm p}<t<2t_{\rm p}$, 
where $\bi{B}_\pm = \brkt{\bi{B}_0 \pm g^{-1}\om \hat{\bi{z}}}$.
Having removed the time-dependent, we calculate the evolution of the system 
in the frame which has its z-axis parallel to the field 
(either $\bi{B}_+$ or $\bi{B}_-$) these frames we call the  
{\it plus}- and {\it minus}-basis respectively 
(the former is shown in  Fig. \ref{fig1}). 
Finally we rotate back to the lab-frame to evaluate observables.

Before we give a detailled explanation of how we calculate the 
spin's evolution in the presence of the environment, we giving our results.
The anisotropic nature of the coupling results in
$P(\xi)$ containing ${\cal O}[\tilde{C}^2]$-terms 
which go like $\exp[\pm \rmi gBt_{\rm p}]$.
To simplify the resulting expressions 
we average $t_{\rm p}$ over a range $\gtrsim (gB)^{-1}$ 
to remove these terms, then
\vskip -0.20truein
\begin{eqnarray}
P(\xi) &=& \half 
\Big[ 1 + 
\e^{-2t_{\rm p}/T_2}
\cos\brkt{\xi - 4\Phi_{\rm BP}- \kappa_1} \nonumber \\
& & +\stbrkt{\kappa_2}\e^{-2t_{\rm p}/T_2} 
\cos \brkt{\xi + 4\Phi_{\rm BP}- \arg [\kappa_2]}
\nonumber \\
& & +\stbrkt{\kappa_3} \big({2\e^{-2t_{\rm p}/T_2} -\e^{-4t_{\rm p}/T_2}}\big)
\cos \brkt{\xi - \arg \sqbrkt{\kappa_3}} \nonumber \\
& & -\kappa_4 \cos \xi \Big]\ ,
\label{eq:results-10a}
\end{eqnarray}
where $\Phi_{\rm BP} = \Phi^{(0)}_{\rm BP} + \de\Phi_{\rm BP}$.
For compactness we have dropped an uninteresting 
real ${\cal O}[\tilde{C}^2t_{\rm p}^0]$ 
term from the first exponent while retaining such terms elsewhere.
The $\kappa$s (which were schematically refered to as $\Phi_{\rm shift}$ 
above) are ${\cal O}[\tilde{C}^2]$ and so are comparable to 
$\de \Phi_{\rm BP}$; however they are independent
of the time-dependence of ${\cal H}(t)$, and hence independent of
the winding number, $n$.
We find, 
\begin{eqnarray}
T_2^{-1} &=&  (2T_1)^{-1}
= {\textstyle {\pi \over 8}} \tilde{C}^2 \Om_{\rm m} \gamma \e^{-\gamma} 
\sin^2 \theta 
\label{eq:results-20a} \\
\de \Phi_{\rm BP} 
&=&   {\textstyle {\pi \over 8}} n \tilde{C}^2 
\sqbrkt{f'(\gamma) -2\gamma^{-1}f(\gamma) }\sin^2 \theta \cos \theta \ ,
\label{eq:results-20b}
\end{eqnarray} 
where $\gamma=gB/\Om_{\rm m}$. 
The function $f(x)=x\e^x {\rm Ei}(-x) + x\e^{-x}{\rm Ei}(x)$
where we define ${\rm Ei}(x)$ as the principal-value of the
Exponential integral, $\int_{-x}^\infty \rmd t \e^{-t}/t$, and
$f'(x)\equiv{\rmd f(x)/\rmd x}$.
Eq. (\ref{eq:results-20b}) is simply the $\om$-dependent term
in the Lamb shift of the energy when in the rotating frame.  
This generates a term of  ${\cal O}[n \! \cdot \! t_{\rm p}^0]$ 
in the phase which in the laboratory frame is a contribution to the BP.

The $n$-independent factors are 
\vskip -0.25truein
\begin{eqnarray}
\kappa_1 
&=&{\textstyle{1 \over 4}}\tilde{C}^2 \pi \e^{-\gamma} \sin \theta 
\sqbrkt{\cos \theta + {\textstyle{1 \over 4}}\sin \theta}
\nonumber \\
\kappa_2 
&=& {\textstyle{1\over 16}}\tilde{C}^2 \brkt{\gamma^{-1} f(\gamma) 
+\rmi \pi \e^{-\gamma}}\sin^2 \theta
\nonumber \\
\kappa_3 
&=& {\textstyle{1\over 4}}\tilde{C}^2 
\sqbrkt{ \brkt{\gamma^{-1} f(\gamma) +\rmi \pi \e^{-\gamma} }
\sin\theta \cos\theta  - 2\gamma^{-1} \sin \theta }
\nonumber \\
\kappa_4 
&=&{\textstyle{1\over 2}}\tilde{C}^2 
\sqbrkt{\e^\gamma {\rm Ei}(-\gamma)
\sin \theta \cos \theta 
-\gamma^{-1} \sin \theta }\ .
\label{eq:results-30}
\end{eqnarray} 
Now we check that the BP 
is not simply given by the solid-angle enclosed by the spin-resonances.
{\em If} this were the  case
then the BP for this experiment would be
$\Phi^{(0)}_{\rm BP} 
- {\textstyle{\pi \over 4}} 
n \tilde{C}^2 \gamma^{-1}f(\gamma) \sin^2\theta \cos\theta$,
the correct result deviates from this prediction by
\vskip -0.25truein
\begin{eqnarray}
\de \Phi'_{\rm BP}&=& 
{\textstyle{\pi \over 8}} n \tilde{C}^2 
f'(\gamma) \sin^2 \theta  \cos\theta
\label{eq:results-40} \ ,
\end{eqnarray}
for most $\gamma$ this deviation is significant.


We now discuss the method we use to obtain 
these results.
The Hamiltonian in the primed-basis 
is time-independent and is given by 
\vskip -0.25truein
\begin{eqnarray}
{\cal H}'(\bi{B}_\pm) &=& 
-{g \over 2} \bi{B}_\pm \cdot \hat{\bi{\sigma}}
+\sum_{j} \Om_j \brkt{\hat{b}^\dagger_{j}\hat{b}_{j} + \half}
\nonumber \\
& & \quad \quad -{g \over 2}\sum_{j}{C\over (2m\Om_j)^{1/2}} 
\brkt{\hat{b}^\dagger_{j}+\hat{b}_{j}}\hat{\sigma}_z \ . \quad 
\label{eq:details-20}
\end{eqnarray} 
%
If we write the spin's initial density matrix as ${\bf \rho}_0$ and the 
oscillators initial density matrix as  $\rho^{\rm osc}_0$, then we are 
interested in the spin density matrix at time $t$,
after we have traced over the oscillator states, 
$\rho_t 
= \tr_{\rm osc} \hat{U}_t 
\brkt{\rho_0 \otimes \rho^{\rm osc}_0} \hat{U}^\dagger_t$,
where $\hat{U}_t$ is the evolution operator.
We find it helpful to write the spin density matrix as a vector $\bi{\rho}$ 
whose elements are $(\rho_{11}, \rho_{12},\rho_{21}, \rho_{22})$. 
Then the spin evolution equation (after the oscillators have been traced over) 
can be written as
$\bi{\rho}_t = {\bf K}(t) \bi{\rho}_0$,
where this defines ${\bf K}(t)$ as a four-by-four matrix which 
gives the time evolution of the elements of the spin's density matrix.
The initial state of the oscillators enters in the functional form of 
the elements of ${\bf K}(t)$. 
For the experiment described above eq. \eref{eq:intro-10} we need to
calculate  
$\bi{\rho}_{2t_{\rm p}}
= {\bf K}^{\rm flip} \ {\bf K}(\bi{B}_+,t_{\rm p})\  
{\bf K}^{\rm flip}\ 
{\bf K}(\bi{B}_-,t_{\rm p})\ 
\bi{\rho}_0$.
The spin-flip is assumed to be fast enough to leave all the oscillators 
unchanged while flipping the spin, then in the primed-basis
${\bf K}^{\rm flip}$ simply has ``1''s on the off-diagonal and ``0''s 
elsewhere.
This leaves the calculation of the propagation matrix 
${\bf K}(\bi{B}_+,t_{\rm p})$, we can find 
${\bf K}(\bi{B}_-,t_{\rm p})$ by reversing 
the sign of $\om$ throughout.
For weak coupling to the bath it is natural to work
in the {\em plus-basis} (see Fig.~\ref{fig1}), 
which has its $z$-axis parallel to $\bi{B}_+$, 
in this basis ${\bf K}(\bi{B}_+,t_{\rm p})$ becomes diagonal 
if $\tilde{C} \to 0$.
Finally the coupling between spin and oscillators in the plus-basis
at time $t$ is ${\bf C}_+(t) = {\bf C}'(t){\bf R}_+$, 
where ${\bf R}_+$ is the SO(3) rotation from the primed-basis 
to the plus-basis. 

\begin{figure}[t]
  \includegraphics[scale=0.55]{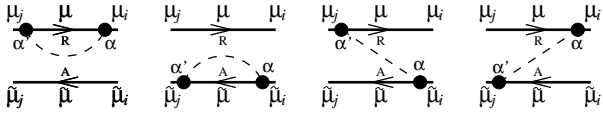} 
  \caption{
  The four classes of $\ord{\tilde{C}^2}$-contributions to $\Sigma^+(\tau)$ 
  are shown here.  The upper (lower) line is the Retarded (Advanced) 
  non-interacting ($C_\al=0$) spin-propagator.
  Everything is in the basis where these propagators are diagonal
  (the plus-basis for $\Sigma^+$).
  The spin-state is $\mu$ with $\up (\dn) \equiv +1(-1)$.
  The dotted-lines are the spin-spin interactions after we have traced out the 
  oscillators which couple to the spin via $\sigma_\al$.}
\label{fig2}
\vskip -0.2truein
\end{figure}

Now we use the real-time transport method \cite{Schoeller94}
to write the following differential equation for  ${\bf K}^+(t)$,
\vskip -0.25truein
\begin{eqnarray}
\partial_t {\bf K}^+(t) = -\rmi {\bf E}^+ {\bf K}^+(t) 
+ \int_0^t \rmd \tau {\bf \Sigma}^+(\tau){\bf K}^+(t-\tau)  \ , 
\label{eq:details-30}
\end{eqnarray}
where all bold symbols are $4\times 4$ matrices.
The matrix ${\bf E}^+$ gives the evolution of the propagation matrix 
when there is no coupling to the bath.  Because we are in the plus-basis
it is diagonal with 
$E^+_{11}=E^+_{44}=0$ and 
$E^+_{22}=-E^+_{33}=-gB^+\equiv -g\stbrkt{\bi{B}+g^{-1}\bi{\om}}$.
The matrix ${\bf \Sigma}^+(\tau)$ is the contribution of all irreducible 
diagrams with one or more interactions with the bath of oscillators.
Equation \eref{eq:details-30} is exact, however to proceed we
treat this equation to first order in $\tilde{C}^2$.
Thus in the integral on the left hand side of  \eref{eq:details-30}
we treat ${\bf \Sigma}^+(\tau)$ 
to first order in $\tilde{C}^2$ and ${\bf K}^+(t-\tau)$ to zeroth order.
So we can write 
${\bf K}^+(t-\tau) 
\simeq  {\bf K}^+(t) {\bf K}_0^+(-\tau)$
where the corrections to the approximations are ${\cal O}[\tilde{C}^2]$ 
and so can be ignored.  
Now ${\bf \Sigma}^+$, which is evaluated below, is dominated by small-$\tau$ 
so we take the upper limit on the integral to infinity.  
The error we make in doing so is $\ord{(\Om_{\rm m}t_{\rm p})^{-1}}$ 
which we neglect.
This systematic approximation results in the interaction becoming 
local in time.
Then we get
$\partial_t {\bf K}^+(t) = \sqbrkt{-\rmi {\bf E}^+ + {\bf X}^+}{\bf K}^+(t)$
where
${\bf X}^+=\int_0^\infty \rmd \tau {\bf \Sigma}^+(\tau){\bf K}_0^+(-\tau)$,
and diagonalise the matrix
$\brkt{-\rmi {\bf E}^+ + {\bf X}^+}$ to find ${\bf K}^+(t)$.

Now we briefly discuss the evaluation of ${\bf \Sigma}^+$ to lowest order in 
$\tilde{C}^2$.  
At this order we need only consider irreducible diagrams with a single
interaction with the oscillators.
When the oscillators are traced out they leave an 
interaction between the spin at time $t$ and time $t-\tau$.
The resulting first-order irreducible diagrams are shown in Fig.~\ref{fig2}.
The contribution to ${\bf \Sigma}^+$ of the diagram with an interaction via 
$\sigma_\al$ at time $t$ and another via $\sigma_{\al'}$ at time $t'$,
after we have summed over the Ohmic bath, is
\begin{eqnarray} 
& &-\chi {\textstyle {g^2 \al \over 8m}}
\sqbrkt{{\bf C}^{\rm T}_+(t) {\bf C}_+(t')}_{\al\al'}
 \nonumber \\
& &\qquad \times \cbrkt{\sqbrkt{\sigma_\al}_{\mu_i\mu} 
\atop \sqbrkt{\sigma_\al}_{\tilde{\mu}\tilde{\mu}_i}}
\cbrkt{\sqbrkt{\sigma_{\al'}}_{\mu \mu_j} 
\atop \sqbrkt{\sigma_{\al'}}_{\tilde{\mu}_j\tilde{\mu}}}
{\Om_{\rm m}^2 \e^{\rmi gB^+ (\mu-\tilde{\mu})\tau/2}
\over(1+ \rmi \kappa \Om_{\rm m} \tau)^2} \ ,\qquad \quad
\end{eqnarray}
where $\kappa=\pm1$, $\chi =\pm 1$,
and other variables are shown in Fig.~\ref{fig2}.
The upper (lower) 
term in $\cbrkt{\cdots}$ is applicable if the relevant vertex is
on R (A). 
$\kappa$ is $+1$ ($-1$) when the $\al'$ vertex is on R (A).
$\chi$ is $+1$ ($-1$) if the interaction is R--R or A--A 
(R--A or A--R).

In conclusion,
the BP can be observed in a non-isolated system,
if the coupling to the environment is weak enough that $gB \gg T_2^{-1}$.
The adiabatic phase is 
$\Phi^{(0)}_{\rm BP} + \de \Phi_{\rm BP} + \Phi_{\rm shift}$,
but $\Phi_{\rm shift}$ is not considered a BP because it does not vanish when
$n=0$.  So the BP 
differs from that of a isolated spin by $\de \Phi_{\rm BP}$, given in 
Eq. (\ref{eq:results-20b}).
The proportionality of $\de \Phi_{\rm BP}$ to $n$ hints that it has some 
geometric character,  however it is a function of the environment's spectrum
and thus the total BP is not a simple geometric quantity.


We are extremely grateful to A. Shnirman for useful discussions,
and we thank R. Fazio, F. Wilhelm and Y. Aharonov for enlightening comments.
This work was commenced while RW was working at the Weizmann Institute
and was supported by the U.S.-Israel Binational Science Foundation
(BSF), by the Minerva Foundation, by the Israel Science Foundation, 
and by the German-Israel Foundation (GIF).



\begin{thebibliography}{10}

\bibitem{Falci00}
G. Falci {\it et al},
{\em Nature} {\bf 407} 355 (2000).

\bibitem{Berry84}
M.V. Berry, {\em Proc. R. Soc. Lond.} {\bf 392} 45 (1984).

\bibitem{Jones00Ekert00}
J.A. Jones {\it et al},
{\em Nature} {\bf 403}, 869 (2000).
A. Ekert {\it et al},
{\em J. Mod. Opt.} {\bf 47},2501, (2000).

\bibitem{Wilhelm01}
F.K. Wilhelm and J.E. Mooij, to be published.

\bibitem{Nakamura99}
Y. Nakamura, Yu. A. Pashkin and J.S. Tsai {\em Nature} {\bf 398} 786 (1999)

\bibitem{Anandan97}
J. Anandan, J. Christian and K. Wanelik,
{\em Am. J. Phys.} {\bf 65} 180 (1997) and references therein.




\bibitem{Gaitan98}
F. Gaitan, {\em Phys. Rev. A} {\bf 58} 1665 (1998).

\bibitem{Ao99}
P. Ao and X.-M. Zhu, {\em Phys. Rev. B} {\bf 60} 6850 (1999).


\bibitem{Avron98Avron99}
J.E. Avron and A. Elgart, {\em Phys. Rev. A} {\bf 58} 4300 (1998);
{\em Comm. Mat. Phys.}{\bf 203} 445 (1999).

\bibitem{endnote-about-Avron}
This implies that the work on adiabatic evolution
of the ground-state of models similar to ours \cite{Avron98Avron99}
is not directly relevant to our work.
 
\bibitem{Leggett87}
A.J. Leggett {\it et al},
{\em Rev. Mod. Phys.} {\bf 59} 1 (1987),
and references therein.


\bibitem{wg-tbp}
The generalisation of our analysis to finite temperatures
will be published elsewhere.


\bibitem{non-adiab}
M.V. Berry {\em Proc. R. Soc. Lond. A} {\bf 414} 31 (1987);
N. Datta, G. Ghosh and M.H. Engineer {\em Phys. Rev. A} {\bf 40} 526 (1989);
F. Gaitan {\em J. Mag. Resonance} {\bf 139} 152 (1999).

\bibitem{Shnirman02}
A. Shnirman, Y. Makhlin, G. Schoen, to be published
-- eprint : cond-mat/0202518.



\bibitem{endnote-isotropic}
We have also considered both $z$-axis ($C_\al=C\de_{\al z}$) 
and isotropic coupling ($C_\al=C$), 
starting from the ground state of the uncoupled system
(turning on the coupling between step (a) and (b)).  
For $z$-axis coupling, the BP is identical to the above results,
but for isotropic coupling the BP is unmodified by the environment.


\bibitem{Schoeller94}
H. Schoeller and G. Schoen
{\em Phys. Rev. B} {\bf 50} 18436 (1994)

 

\end{thebibliography}
\end{document}